\documentclass[a4paper]{jpconf}
\usepackage{graphicx}
\usepackage{amsfonts,amssymb,amsthm}
\usepackage{psfrag}
\usepackage{array}
\usepackage{bbm}
\newcommand{\f}{\frac}

\newcommand{\su}{\mathfrak{su}}
\newcommand{\an}{\mathfrak{an}}
\newcommand{\so}{\mathfrak{so}}
\newcommand{\SU}{\mathrm{SU}}
\newcommand{\AN}{\mathrm{AN}}

\newcommand{\SO}{\mathrm{SO}}

\newcommand{\R}{\mathbb{R}}

\newcommand{\N}{\mathbb{N}}
\newcommand{\Z}{\mathbb{Z}}

\newcommand{\cM}{{\cal M}}

\newcommand{\uu}{{\cal U}}

\newcommand{\tF}{{\tilde F}}

\newcommand{\id}{\mathbb{I}}

\newcommand{\be}{\begin{equation}}
\newcommand{\ee}{\end{equation}}
\newcommand{\bes}{\begin{eqnarray}}
\newcommand{\ees}{\end{eqnarray}}

\def\tl{\widetilde}

\def\eps{\epsilon}

\def\ka{\kappa}
\def\cc{{\cal C}}
\def\la{\langle}
\def\ra{\rangle}

\renewcommand{\hat}{\widehat}
\def\kk{{\cal K}}

\def\ss{{\cal S}}
\def\tlF{\tl{F}}

\def\mn{{\mu\nu}}

\def\nn{\nonumber}
\begin{document}
\title{Emergent non-commutative matter fields from Group Field Theory models of quantum spacetime}

\author{Daniele Oriti}

\address{Institute for Theoretical Physics, Utrecht University, \\
Leuvenlaan 4, 3584 TD, Utrecht, The Netherlands, EU \\ and \\
Albert Einstein Institute, \\ Am Muehlenberg 4, D-14476 Golm, Germany, EU}

\ead{daniele.oriti@aei.mpg.de}

\begin{abstract}
We offer a perspective on some recent results
obtained in the context of the group field
theory approach to quantum gravity,
on top of reviewing them briefly. These concern a natural
mechanism for the emergence of non-commutative field theories for
matter directly from the GFT action, in both 3 and 4 dimensions
and in both Riemannian and Lorentzian signatures. As such they represent an important step, we argue, in bridging the gap
between a quantum, discrete picture of a
pre-geometric spacetime and the effective continuum geometric
physics of gravity and matter, using ideas and tools from field theory and condensed matter analog gravity models, applied directly at the GFT level. \end{abstract}

\section{Introduction}
In this contribution we offer a perspective on some recent results
\cite{eterawinston,noi} obtained in the context of the group field
theory (GFT) approach to quantum gravity \cite{iogft,laurentgft},
on top of reviewing them briefly. These concern a natural
mechanism for the emergence of non-commutative field theories for
matter directly from the GFT action, in both 3 and 4 dimensions
and in both Riemannian and Lorentzian signatures. The interest of
such results is manifold. First, they show a straightforward link
between spin foam/loop quantum gravity models, via GFTs, and
non-commutative geometry. Second, they are, in principle, a
crucial step forward in the attempt to relate this class of models
with effective frameworks, like Deformed Special Relativity, that form the basis of much current Quantum Gravity
Phenomenology \cite{QGPhen}. On top of all this, in this contribution
we argue that they can be naturally understood within a certain
interpretative framework for GFTs and for the relation between its quantum microscopic spacetime
structures and continuum spacetime physics. The perspective we
present is not, of course, the only possible consistent one, nor
the one shared by all the researchers working in this specific
area. It is, however, a scenario that is consistent with what
we presently know about GFTs, and with a recent proposal about the
emergence of continuum physics from them, as put forward by the
author. Moreover, it is a scenario that makes these results all
the more exciting, which is, we believe, an added benefit. It is based on three main points: 1) GFTs are the most
complete and fundamental definition of quantum gravity models
based on spin network or simplicial gravity structures; 2)
continuum spacetime physics is to be looked for in the collective
\lq\lq many-particle\rq\rq physics of GFT quanta; 3) therefore,
instead of usual LQG, spin foam or simplicial gravity techniques,
one could try to adopt a condensed matter perspective on quantum
spacetime, and see what condensed matter ideas and tools give when
applied to the quantum spacetime system directly at the GFT level; 4) in particular, this shift in
perspective and tools can be very useful for bridging the gap
between a microscopic, quantum, discrete picture of a
pre-geometric spacetime and the effective continuum geometric
physics of gravity and matter we are accustomed and have phenomenological access to. We have exposed the above points and drawn some
possible consequences of this perspective in \cite{gftfluid}. Here
we point out how the results on the emergence of non-commutative
field theories from GFTs can fit in it.

\subsection{Group Field Theory formulation of Quantum Gravity}
Group Field Theories \cite{iogft, laurentgft} are quantum field
theories on (possibly, extensions of) group manifolds
characterized by a peculiar non-local coupling of fields, designed
to produce, in their perturbative expansion around the vacuum,
Feynman diagrams that can be put in 1
to 1 correspondence with d-dimensional simplicial complexes. The
fundamental field is interpreted as a second quantized
(d-1)-simplex or as a (second quantized) elementary spin network
vertex. The Feynman amplitudes can be understood as simplicial
gravity path integrals (although not necessarily written in terms
of some gravity action) or, dually, as spin foam models, i.e. sum
over histories of spin networks states. GFTs are thus at the same time
a generalization of matrix models for 2d quantum gravity to higher
dimensions, and a new version of the simplicial quantum gravity
approaches, and a complete definition of the covariant dynamics of
spin networks in loop quantum gravity.
Let us be a bit more specific.

In their simplest type of models, the field is a
$\mathbb{C}$-valued function of d group elements
$\phi(g_1,..,g_d)$, for a group $G$ (configuration space) being
for example the $SO(d-1,1)$ Lorentz group. Each argument corresponds to one of the
boundary (d-2)-faces of the (d-1)-simplex represented by the
field. Additional symmetry requirements can be imposed. Typically,
one imposes invariance under diagonal action of $G$:
$\phi(g_1,...,g_d)=\phi(g_1g,...,g_dg)$. Fields and action can then be expanded in modes, i.e. in group representations. Both
group and representation variables can be given a
straightforward interpretation (justified by their role in the
corresponding amplitudes) as (pre-) geometric data, i.e. data
defining a discrete geometry for the simplicial complex associated
with each Feynman diagram. The details of course depend on the specific
model. A GFT model is defined by a choice of action:
$$
S= \frac{1}{2}\int
  dg_id\tilde{g}_i\,
  \phi(g_i)\mathcal{K}(g_i\tilde{g}_i^{-1})\phi(\tilde{g}_i)
  +
  \frac{\lambda}{(d+1)!}\int dg_{ij}\,
  \phi(g_{1j})...\phi(g_{d+1 j})\,\mathcal{V}(g_{ij}g_{ji}^{-1}),
$$
i.e. of a kinetic and interaction functions $\mathcal{K}$ and
$\mathcal{V}$. The interaction term describes the interaction of (d-1)-simplices to form a d-simplex by gluing them along their
(d-2)-faces (arguments of the fields); this gives the mentioned non-local
combinatorial pairing of fields. The nature of
the interaction is specified by the choice of function
$\mathcal{V}$. The kinetic term involves two fields each
representing a given (d-1)-simplex seen from one of the two
d-simplices (interaction vertices) sharing it, so that the choice
of kinetic functions $\mathcal{K}$ specifies how the geometric
degrees of freedom corresponding to their d (d-2)-faces are
propagated from one vertex to the next. The quantum theory is
defined in terms of the expansion in Feynman diagrams of the
partition function:

$$ Z\,=\,\int
\mathcal{D}\phi\,e^{-S[\phi]}\,=\,\sum_{\Gamma}\,\frac{\lambda^N}{sym[\Gamma]}\,Z(\Gamma),
$$
where $N$ is the number of interaction vertices in the Feynman
graph $\Gamma$, $sym[\Gamma]$ is a symmetry factor for the graph
and $Z(\Gamma)$ the corresponding Feynman amplitude. By
construction (\cite{iogft,laurentgft} the Feynman
diagram $\Gamma$ is a collection of 2-cells (faces), edges and
vertices, i.e. a 2-complex, topologically dual to a d-dimensional
simplicial complex. The resulting complexes/triangulations can
have arbitrary topology, each corresponding to a particular {\it
interaction process} of the fundamental building blocks of space,
i.e. (D-1)-simplices. In representation space each Feynman diagram
is a spin foam (a 2-complex with faces $f$ labelled by group
representations), and each Feynman amplitude defines a spin foam model.
Because of the geometric interpretation for the (group-theoretic)
variables these Feynman amplitudes define a sum-over-histories
for discrete quantum gravity, i.e. a sum over simplicial geometries, within a sum over
simplicial topologies. 

Several GFT models have been and are currently studied, and much
more is known about GFTs in general. For all this, we refer to the
literature \cite{iogft,laurentgft,lqg,alex}.

\subsection{From GFT to continuum spacetime physics?}
GFTs are thus a (tentative) class of models for quantum
spacetime, and thus {\it not defined on any spacetime}, in which
its microscopic structure is described by elementary building
blocks given by spin network vertices or simplices, labelled by
pre-geometric data. This description is valid and useful in the
regime in which the perturbative expansion of the GFT partition
function is valid, i.e. small coupling constant and few GFT quanta
involved in the physical process one is considering. In this
approximation a discrete spacetime {\it emerges}
as a Feynman diagram from this \lq\lq pre-spacetime\rq\rq theory.

This is one of the intriguing aspects of GFTs. It is also the
origin of the main open issue: how to recover a continuum
spacetime, its continuum geometry and all continuum
physics, including usual General Relativity and quantum field
theories for matter, from such a different
description of the pre-geometric regime? This is the
outstanding problem faced by {\it all} current discrete approaches
to quantum gravity\cite{libro}.

The problem of the continuum involves actually several intertwined
questions. Some of them are conceptual: where does a concept of
geometry originate from in a non-geometric or pre-geometric
framework? where to look for notions and properties of space and
time in models defined in absence of any spacetime structure? Some
are purely kinematical: should we expect the pre-geometric GFT
data to give rise to the continuum geometry? or could an emergent
macroscopic continuum geometry be totally unrelated to these
microscopic and pre-geometric data? what is the correct
approximation scheme to obtain a continuum spacetime from the
discrete GFT structures? Some are instead dynamical: what is the
physical regime, the physical conditions, in which the continuum
approximation is viable, justified and/or useful? what is the
physical dynamical process leading to this regime from the
pre-geometric one, and how to describe it?

We have discussed in \cite{gftfluid} our personal perspective on
this problem, and the possibility that GFTs could be the right
setting for tackling it, and how.
We summarize it here, briefly. From the GFT point of view, the
crucial issue is whether we expect the continuum approximation (as
opposed to large scale, semi-classical or other a priori distinct
approximations) to involve very large numbers of GFT quanta or
not\footnote{In the full theory; the same question in a symmetry
reduced context, for example, may have a different answer.}. We
opt for a positive answer, as naive reasoning would suggest (one
would expect a generic continuum spacetime to be formed by
zillions of Planck size building blocks, rather than few
macroscopic ones). If this is
the case, then we are dealing, from the GFT point of view, with a
many-particle system whose microscopic theory is given by some
fundamental GFT action and we are interested in its
collective dynamics and states in some thermodynamic
approximation. This simple thought alone suggests us to look for
ideas and techniques from statistical field theory and condensed
matter theory, and to try to apply/reformulate/re-interpret them
in a GFT context.This also immediately suggests that the GFT formalism is the most
natural setting for studying this dynamics even coming from the
pure loop quantum gravity perspective or from simplicial quantum
gravity. In the first case, in fact, GFTs offer a second quantized
formalism for the same quantum geometric structures, and quantum
field theory is indeed what comes natural in condensed matter when
dealing with large collections of particles/atoms. In the second
case, GFTs offer an alternative non-perturbative definition of the
quantum dynamics of simplicial gravity, re-interprets the usual
sum over discrete geometries as a perturbative expansion around
the no-spacetime vacuum, and in doing so suggest the possibility
of different vacuum states and a different reformulation of the
same dynamics that is better suited for studying the dynamics of
(combinatorially) complicated simplicial geometries. See
\cite{gftfluid} for more details.

\section{Condensed matter perspective and GFT implementation}
So, what can we learn from condensed matter models in our quest
for bridging the gap between microscopic quantum and discrete
pre-geometry in absence of spacetime, and macroscopic continuum
spacetime and associated physics, with matter fields,
geometry, gauge interactions and all that?

Condensed matter theory comes in help in two main ways, one very
general, one more specific.

The first: condensed matter theory is exactly the area of physics
concerned with the general issue of understanding the physical
behaviour of large assemblies of microscopic constituents, in
their collective behaviour, of elucidating their emergent
properties, and of developing the appropriate notions and
mathematical tools for describing the different layers of
collective organization that coexists in a given phase, as well as
the transition from one phase to another.
Therefore, once we realize that a classical continuum spacetime is the result
of the collective properties and dynamics of large numbers of GFT
quanta,  it is natural to look at condensed matter
theory for insights.

One could be a bit more specific (and speculative) and conjecture
that continuum spacetime arises as a {\it fluid phase} of this
large assembly of GFT quanta, maybe after a process
analogous (mathematically) to condensation in GFT momentum space
(Bose condensation, for example). This conjecture is also argued
for in \cite{gftfluid}. Continuum geometrodynamics, be it
classical (General Relativity) or quantum could
then arise from the effective GFT hydrodynamics in this fluid (or
condensed) phase. This is a second conjecture put forward in
\cite{gftfluid}. Again, in order to put these ideas to test,
one has to turn to condensed matter theory and to statistical
field theory, i.e. we should develop statistical group
field theory, and study that special condensed
matter system that is a quantum spacetime. A corresponding research programme has been lied down tentatively
in \cite{gftfluid}, and it involves developing
first for GFTs and then applying several techniques, common in statistical
field theory and condensed matter. One of them is the
renormalization group, and in fact one way to study the phase
structure of GFTs as well as the behaviour of GFT systems at
different scales would certainly be in terms of perturbative and
non-perturbative renormalization group transformations. No
systematic study of GFT renormalization has been performed until
now, but work is in progress from various groups. In particular, a
systematic treatment of perturbative GFT renormalization for
the 3d Boulatov model is under way, and its first results can be found in
\cite{renormBoulatov}. Another avenue would be to develop an
Hamiltonian statistical mechanics for GFTs, and in turn this
involves several steps: a canonical/hamiltonian GFT formalism,
which made highly non-trivial by the non-local nature of these
models, a precise understanding of the GFT analogue of the notion
of energy, temperature, pressure, etc, and a re-interpretation of
any such notion in quantum gravity terms. With
these tools at hand one should analyze the phase structure of
various GFT models and prove the existence of an appropriate fluid
phase in some of them. Then, the hydrodynamic approximation in
this phase should be developed and the final goal would be to show
the emergence of General Relativity from this GFT hydrodynamic
description. It is
too early to report on progress on this front.

There is a second way, however, in which condensed matter theory
comes into the game. It provides specific examples of systems in which the
collective behaviour of the microscopic constituents in some
hydrodynamic approximation gives rise to effective emergent
geometries as well as matter fields. Thus it gives
further support, by means of explicit examples, to the idea that
continuum geometry and gravity may emerge naturally from
fundamental systems which do not have a geometric or gravitational
nature per se, at least in their fundamental formulation (GFTs are
of this type). These are the so-called analog gravity models
\cite{analog}.

The emergence of gravity and (generically) curved geometries in
analog condensed matter models takes place in the hydrodynamic regime, i.e.
usually at the level of a (modified) Gross-Pitaevskii equation or
of the corresponding Lagrangian/Hamiltonian, which in turn is
usually obtained in the mean field theory approximation (or
refined version of the same) around some background configuration
of the fluid under study. The background configurations that have
proven more interesting are those identifying Bose-Einstein
condensates or fermionic superfluids \cite{analog}. However, the
emergence of effective metric fields is more general and not
confined to these rather peculiar systems. What happens is that
the collective parameters describing the fluid and its dynamics in
these background configurations (e.g. the density and velocity of
the fluid in the laboratory frame) can be recast as the component
functions of an {\it effective metric field}.

This would be only cosmetics if not for two further results. 1)
In some very special cases and in some particular approximation
the hydrodynamic equations governing the dynamic of the effective
metrics, when recast in geometric terms, can also be seen to
reproduce known geometrodynamic theories, at least in part,
ranging from Newtonian gravity to (almost) GR; see
\cite{analog2}. This means that it is possible to reproduce also
gravitational dynamics as emergent from systems that are not
geometric in nature or form. 2) The
effective dynamics of perturbations (quasi-particles, themselves
collective excitations of the fundamental constituents of the
fluid) around the same background configurations turns out to be
given by matter field theories in curved spacetimes, whose
geometry is indeed the one identified by the effective metrics
obtained from the collective background parameters of the fluid.

The first type of results is so far limited to special systems,
peculiar approximations, and ultimately not fully satisfactory, in the sense that it has not been possible yet to
reproduce, say, the Einstein-Hilbert dynamics
in any, however idealized, condensed matter system \cite{analog2}.
This seems to suggest that the ideas and techniques developed in
this context and to this aim are certainly useful and interesting,
but need to be applied to some very peculiar system representing
(quantum) spacetime, if one wants to explain in this way the
emergence of geometry and General Relativity in the real world.
Therefore, these results certainly encourage and support the idea
of applying similar ideas and techniques in Group Field Theory, but they do not
provide a sure guidance on how exactly this should be done.

The second type of results, however, is much more general and
applies to a very large class of systems and approximations,
including systems as common as ordinary fluids (e.g. water) in
everyday physical conditions \cite{analog}. It is this type of
results that we focus on in this contribution, because we are able to obtain similar results in a GFT
context.

The general scheme of what goes on in all these condensed matter
systems (including BEC, superfluids, etc) concerning the emergence
of matter field theories on effective metric spacetimes is well
captured by the following \lq\lq meta-model\rq\rq described in
\cite{analog}. Consider a system described by a single scalar
field $\phi(x)$, living on a flat metric spacetime of trivial
topology (a good approximation for the quantum geometry of any lab
in any research institute on Earth). The field $\phi$ can be the
effective order parameter for a BEC, the collective field encoding
the velocity and density of an ordinary fluid in the hydrodynamic
approximation, or whatever else. Assume that its dynamics is encoded in a lagrangian
$L(\phi,\partial_\mu\phi)$ depending on the field and its partial
derivatives. Let us expand the field around some classical
solution $\phi_0$ of the equations of motion, as:
$\phi(x)=\phi_0(x)+\phi_1(x)$. Next we expand the lagrangian
itself to obtain an effective action for the perturbation field
$\phi_1(x)$ (we focus our attention to the kinetic term only); we
obtain, generically, an effective Klein-Gordon operator on a
curved metric: \bes
S(\phi)\,&=&\,S(\phi_0)\,+\,\frac{1}{2}\,\int\,\left[
\frac{\partial^2
L}{\partial(\partial_\mu\phi)\partial(\partial_\nu\phi)}\mid_{\phi_0}\,\partial_\mu\phi_1\partial_\nu\phi_1\right.
\left.+\,( \frac{\partial^2L}{\partial\phi\partial\phi}\,-
\,\partial_\mu\frac{\partial^2L}{\partial(\partial_\mu\phi)\partial\phi})\mid_{\phi_0}
\right]\,+\, (.....) \,=\, \nonumber \\
&=&\, S(\phi_0)\,+\, \frac{1}{2}\int\,\sqrt{-g}\left[ \phi_1
\square_{\phi_0}\,\phi_1\,-\,V(\phi_0)\,\phi_1^2 \right]+\,
(.....\textit{interactions}),\ees with the operator
$\square_{\phi_0}= g^{\mu\nu}\partial_\mu\partial_\nu$, for the
effective (inverse) metric $\sqrt{-g}\,
g^{\mu\nu}=\frac{\partial^2
L}{\partial(\partial_\mu\phi)\partial(\partial_\nu\phi)}\mid_{\phi_0}$.
It is the possible to invert to obtain the effective metric and
from this the other tensors characterizing the effective spacetime
geometry. Notice that both the \lq\lq fundamental\rq\rq field and
the quasi-particle one live on a 4-dimensional spacetime of
trivial topology, although endowed with a different metric in general. We refer to the literature for further details and
applications of the above general result. The main point to notice
here is that one generically obtains an effective spacetime
geometry to which the quasi-particles couple, depending only in
its precise functional form on the fundamental Lagrangian $L$ and
on the classical solution $\phi_0$ chosen; they do not
couple to the initial (laboratory) flat background metric.

This is the type of mechanism we want to reproduce in a Group
Field Theory context. Assuming that a given GFT model (Lagrangian) describes
the microscopic dynamics of a
{\it discrete quantum spacetime}, and that some solution of the
corresponding fundamental equations can be interpreted as
identifying a given quantum spacetime configuration, 1) can we
obtain an effective macroscopic {\it continuum} field theory for
matter fields from it? and if so, 2) what is the effective
spacetime and geometry that these emergent matter fields see?

Answering these questions means, as we have tried to emphasize,
making an important step towards bridging the gap between our
fundamental discrete models of spacetime, and the usual
continuum description of spacetime.
It also means getting closer to possible quantum gravity
phenomenology, and to experimental falsifiability. Let us also notice that, while the
correspondence between classical solutions of the fundamental
equations and effective geometries is a priori unexpected in
condensed matter systems, which are non-geometric in nature, so that they are referred to as {\it analog} gravity models, in the GFT case the situation is different. We have
here models which are non-geometric and far from usual
geometrodynamics in their formalism, but which at the same time
are expected to encode quantum geometric information and indeed to
determine, in particular in their classical solutions, a (quantum
and therefore classical) geometry for spacetime \cite{iogft}, also
at the continuum level, the issue being how they exactly do so. We
are, in other words, far beyond a pure analogy. What we find
is that it is possible to apply the same procedure to GFT
models and that one can obtain rather straightforwardly effective
continuum field theories for matter fields. The effective matter
field theories that we obtain most easily from GFTs are quantum field
theories on non-commutative spaces of Lie algebra type.

\section{Results on emergent effective matter}
Let us now introduce and review briefly the results obtained so
far.

\subsection{Non-commutative field theories as group field theories}
First of all we introduce the relevant class of non-commutative field
theories. The basic point is the duality between Lie algebra and corresponding Lie group re-interpreted as the non-commutative version of the usual duality between coordinate and momentum space. More precisely, if we have a non-commutative spacetime of Lie algebra type $[X_\mu, X_\nu] = C_{\mu\nu}^\lambda X_\lambda$, the corresponding momentum space is naturally identified with the corresponding Lie group, on which non-commutative coordinates $X_\mu$ act as (Lie) derivatives. From this perspective, we understand the origin of the spacetime non-commutativity to be the curvature of the corresponding momentum space, a sort of Planck scale \lq\lq co-gravity\rq\rq \cite{kappaM}. The link with GFTs is then obvious: in momentum space the field theory on such non-commutative spacetime will be given, by definition, by some sort of group field theory. The task will then be to derive the relevant field theories for matter from interesting GFT models of quantum spacetime. 

\medskip
In 3 spacetime dimensions the results obtained concern a euclidean non-commutative spacetime given by the $\su(2)$ Lie algebra , i.e. whose spacetime coordinates are identified with the $\su(2)$ generators with $[X_i,X_j]= i \f{1}{\kappa} \epsilon_{ijk} X_k$. Momenta are instead identified with group elements $\SU2)$ \cite{majid}, acquiring a non-commutative addition property following the group composition law. A scalar field theory  in momentum space is then given by a group field theory of the type: 
\be
S[\psi]\,=\,
\f{1}{2}\int_{\SU(2)} dg  \psi(g)\kk(g)\psi(g^{-1})-\f{\lambda}{3!}\int[dg]^3\,
\psi(g_1)\psi(g_2)\psi(g_3)\delta(g_1g_2g_3), 
\ee
in the 3-valent case, where the integration measure is the Haar measure on the group.  The non-commutative Fourier transform \cite{majid} relates then functions on the group and non-commutative functions in the enveloping algebra of $\su(2)$. It is based on the non-commutative plane waves $e_g = e^{i k_i X_i} \in \hat{C}_\kappa(\su(2)) $, where $k_i$ are local coordinates on the group manifold labeling the group element (and thus the momentum) $g$ and bounded as $\mid k\mid \leq \kappa$ (so that it is natural to identify $\kappa$ with a Planckian maximal mass/momentum scale). It reads:
\be
\hat \phi(X)= \int_{\SU(2)} dg \, e_{g(k_i)}\, \phi(g(k)), \quad
X\in \su(2), \quad \hat \phi(X) \in \uu(\su(2)).
\ee
Strictly speaking, this construction works on $\SO(3)$, but can be extended to $\SU(2)$ in several ways \cite{etera}. Using this, the action above can be rewritten in configuration space. We will see below some simple example of such action, derived straightforwardly from a group field theory model for 3d quantum gravity. We notice that one can define a further map from elements of the enveloping algebra of $\su(2)$ to functions on $\mathbb{R}^3$ (isomorphic to the same Lie algebra as a vector space) endowed with a non-commutative star product again reflecting the non-commutative composition of momenta following the rules of group multiplication. See \cite{majid,etera,FKG} for details. We also notice that the Feynman amplitudes of the above scalar field action (with simple kinetic terms) can be derived from the Ponzano-Regge spin foam model coupled to point particles \cite{PR3}, in turn obtainable from an extended GFT formalism \cite{iojimmy}. We will see that the GFT construction to be presented allows to bypass completely the spin foam formulation of the coupled theory.

\medskip

The 4-dimensional non-commutative spacetime that is of most direct relevance for Quantum Gravity phenomenology is so-called $\ka$-Minkowski \cite{kappaM}. We recall here the main features of such space and of the non-commutative field theory defined on it, referring for further details to \cite{noi} and references therein. 
$\ka$-Minkowski space-time can be identified with the Lie algebra $\an_{3}$, which is
a subalgebra of $\so(4,1)$. Indeed, if $J_{\mn}$ are the generators of $\so(4,1)$,  the $\an_{3}$ generators are: \be
X_0=\f{1}{\ka}J_{40},\quad X_k=\f{1}{\ka}(J_{4k}+J_{0k}), \quad
k=1,...,3 \,\, , \ee which, once identified with the coordinates of our non-commutative space, characterize it with the commutation relations: \be
\label{an3} [X_0,X_k]\,=-\f{i}{\ka}X_k,\qquad [X_k,X_l]\,=0, \quad
k,l=1,...,3. \ee
Using this, we can then define non-commutative plane waves with the $\AN_3$ group elements as $h(k_\mu)=h(k_0,k_i)\,\equiv\,e^{ik_0X_0}e^{ik_iX_i}$, thus identifying the coordinates
on the group $k_\mu$ as the wave-vector (in turn related to
the momentum). From here, a non-commutative addition of wave-vectors follows from the group multiplication of the corresponding plane waves.

Crucial for our construction, the Iwasawa decomposition relates $\SO(4,1)$ and $\AN_{3}$ as \cite{klymyk}: \be \SO(4,1)\,=\,
\AN_{3}\,\SO(3,1)\,\cup\,\AN_{3}\cM\,\SO(3,1), \ee where
the two sets are disjoint and $\cM$ is the diagonal
matrix with entries $(-1,1,1,1,-1)$
in the fundamental 5d representation of $\SO(4,1)$. 
Since De Sitter space-time
$dS_{4}$ can be defined as the coset $\SO(4,1)/\SO(3,1)$, an arbitrary point $v$ on it
can be uniquely obtained as: \be v\,=\,(-)^\eps h(k_\mu).v^{(0)}=
h(k_\mu)\cM^\eps.v^{(0)},\qquad \eps=0\textrm{ or }1,\quad
h\in\AN_{3}, \ee where we have taken a reference space-like vector
$v^{(0)}\equiv(0,0,0,1)\in \R^4$, such that its little group is the Lorentz group $\SO(3,1)$ and the action of
$\SO(4,1)$ on it sweeps the whole De Sitter space, and defined the
vector $v\,\equiv\, h(k_\mu).v^{(0)}$ with coordinates:
\be\label{5d momentum}
v_0\,=\, -\sinh\f{k_0}{\ka}+\f{{\bf k}^2}{2\ka^2}e^{k_0/\ka} \;\;\;\;
v_i\,=\, -\f{k_i}{\ka} \;\;\;\;\;
v_4 \,=\, \cosh\f{k_0}\ka -\f{{\bf k}^2}{2\ka^2}e^{k_0/\ka}. \nn
\label{5dparam} \ee The sign $(-)^\eps$ corresponds to the two
components of the Iwasawa decomposition. 
We then introduce the set $\AN^c_{3}\equiv\AN_{3}
\cup\AN_{3}\cM$, such that the Iwasawa decomposition reads
$\SO(4,1)=\AN^c_3\,\SO(3,1)$ and that $\AN^c_3$ is isomorphic to
the full de Sitter space. Actually, one can check that $AN^c_{3}$ is itself a group.
A crucial point is that the component $v_4$ of the above vector is left invariant by the action of the Lorentz group $\SO(3,1)$. This suggests to use this function of the \lq\lq momentum\rq\rq $k_\mu$ as a new (deformed) invariant energy-momentum  (dispersion) relation, in the construction of a deformed version of particle dynamics and field theory on $\ka$-Minkowski spacetime.  This is the basis for much current QG phenomenology \cite{QGPhen}.

Finally, we will need an integration measure on $\AN_3$ in order
to define a Fourier transform. The group $\AN_3$ is provided with
two invariant Haar measures: $\int dh_L=\int d^4k_\mu,\quad
\int dh_R=\int e^{+3k_0/\ka}\,d^4k_\mu$, which are respectively
invariant under the left and right action of the group. The left invariant measure can be obtained from the 5d parametrization used above as: \be\label{measure5d}
\ka^4\int\delta(v_Av^A-1)\theta(v_0+v_4)\,d^5 v_A\,=\,\int
d^4k_\mu\,=\,\int dh_L, \ee so it is the natural measure on $\AN_3$ inherited from the Haar
measure on $\SO(4,1)$. However, this measure is not Lorentz invariant, due to the restriction $v_+ >0$.
To get
a Lorentz invariant measure, we write the same measure as a measure on $\AN^c_3\equiv\AN_3
\cup\AN_3\cM\sim dS$: \be \int dh_L\equiv \int_{\AN_3} dh_L^+ +
\int_{\AN_3\cM} dh_L^-=\int\delta(v_Av^A-1)d^5v. \ee Another way
to obtain a Lorentz invariant measure
is to consider the
elliptic de Sitter space $dS/\Z_2$ where we identify
$v_A \leftrightarrow -v_A$, which amounts to identifying the group
elements $h(k_\mu)\leftrightarrow h(k_\mu)\cM$. This space is
indeed isomorphic to $AN_3$ as a manifold. One way to achieve
nicely this restriction at the field theory level is to consider
only fields on De Sitter space (or on $AN_3^c$) which are however
invariant under the parity transformation $v_A \leftrightarrow
-v_A$ \cite{FKG}. 

For the free real scalar field $\phi: G\rightarrow \R$, we
define the action \be \label{action DSR 1} \ss(\phi)= \int
dh \, \phi(h)\, \kk(h)\, \phi(h), \quad \forall h\in G, \ee
where $dh$ is the left invariant
measure. We then interpret $G=\AN_3,\AN_3^c$ as the momentum space. We
demand $\kk(h)$ to be a function on $G$ invariant  under the
Lorentz transformations, which suggests to use some 
function $\kk(h)=f(v_4(h))$.  Two common choices are
 \be \kk_1(h)= (\kappa
^2-\pi_4(h)) -m^2, \quad \kk_2(h)= \kappa ^2-(\pi_4(h))^2-m^2,
\quad \pi_4=\kappa v_4. \ee
The above action is then Lorentz invariant if we choose a Lorentz invariant measure, i.e. $dh_L$ in the case of either generic fields on $\AN_3^c$ or symmetric fields on $\AN_3$.

Finally, the following generalized Fourier transform relates functions on the group
$\cc(G)$ and elements of the enveloping algebra $\uu(\an_3)$, i.e. non-commutative fields on the non-commutative spacetime $\an_3$, i.e. on $\ka$-Minkowski. For $G=\AN^c_3, \AN_3$, respectively:
\begin{eqnarray}
\label{fourier} &&\hat \phi(X)= \int_{\AN_3} dh_L^+ \, h(k_\mu)\,
\phi^+(k)+\int_{\AN_3\cM} dh_L^- \, h(k_\mu)\, \phi^-(k), \;\;\;
X\in \an_3,
\;\;\hat \phi(X) \in \uu(\an_3) \nn \\
&&\hat \phi(X)= \int_{\AN_3} dh_L \, h(k_\mu)\, \phi(k), \quad
X\in \an_3, \;\; \hat \phi(X) \in \uu(\an_3) \nn
\end{eqnarray}
where we used the non-abelian plane-wave $h(k_\mu)$ \cite{majid,FKG}. The
group field theory action on $G$ can now be rewritten as a
non-commutative field theory on $\ka$-Minkowski (in the $\AN_3$ case) \be \label{action DSR 2}
\ss(\phi)= \int dh_L \, \phi(h)\, \kk(h)\, \phi(h) = \int \,
d^4X\, \left(
\partial_\mu\hat\phi(X)
\partial^\mu\hat\phi(X) + m^2 \hat\phi^2(X)\right). \ee The
Poincar\'e symmetries are naturally deformed in order to be
consistent with the non-trivial commutation relations of the
$\ka$-Minkowski coordinates \cite{FKG}.

\subsection{3d case}
The group field theory we start from is the Boulatov model for 3d quantum gravity \cite{iogft}. We consider a real field $\phi: \SU(2)^3\rightarrow \R$  invariant under the diagonal right action of
$\SU(2)$: \be \phi(g_1,g_2,g_3)=\phi(g_1g,g_2g,g_3g), \quad\forall
g\in\SU(2). \ee The action for this 3d group field theory
involves a trivial propagator and the tetrahedral
vertex: \be S_{3d}[\phi]\,=\, \f12\int[dg]^3
\phi(g_1,g_2,g_3)\phi(g_3,g_2,g_1) -\f{\lambda}{4!}\int [dg]^6
\phi(g_1,g_2,g_3)\phi(g_3,g_4,g_5)\phi(g_5,g_2,g_6)\phi(g_6,g_4,g_1).
\ee The Feynman diagrams of the theory are then, by construction, 3d triangulations, while the corresponding Feynman amplitudes are given by the Ponzano-Regge spin foam model \cite{iogft}.

Now \cite{eterawinston} we look at two-dimensional variations of
the $\phi$-field around classical solutions of the corresponding equations of motion:     

\be \phi(g_3,g_2,g_1)\,=\,\f{\lambda}{3!}\int
dg_4dg_5dg_6 \phi(g_3,g_4,g_5)\phi(g_5,g_2,g_6)\phi(g_6,g_4,g_1).
\ee Calling $\phi^{(0)}$ a generic solution to this equation, we
look at field variations
$\delta\phi(g_1,g_2,g_3)\equiv\psi(g_1g_3^{-1})$ which do not
depend on the group element $g_2$.
We consider a specific class of classical solutions, named ``flat"
solutions (they can be interpreted as quantum flat space on some a priori non-trivial topology): \be
\phi^{(0)}(g_1,g_2,g_3)\,=\,\sqrt{\f{3!}{\lambda}}\,\int dg\;
\delta(g_1g)F(g_2g)\delta(g_3g), \quad F:G\rightarrow\R. \ee As
shown in \cite{eterawinston}, this ansatz gives solutions to the field
equations as soon as $\int F^2=1$.

This leads to an effective action for the 2d variations $\psi$: \be
S_{eff}[\psi]\,=\,
\f12\int\psi(g)\kk(g)\psi(g^{-1})-\f\mu{3!}\int[dg]^3\,
\psi(g_1)\psi(g_2)\psi(g_3)\delta(g_1g_2g_3)
-\f{\lambda}{4!}\int[dg]^4\, \psi(g_1)..\psi(g_4)\delta(g_1..g_4),
\ee with the kinetic term and the 3-valent coupling given in term
of $F$:
$$
\kk(g)\,=\,1-2\left(\int F\right)^2-\int dh F(h)F(hg), \qquad
\f\mu{3!}\,=\,\sqrt{\f{\lambda}{3!}}\,\int F.
$$
with $F(g)$ assumed to be invariant under conjugation $F(g)=F(hgh^{-1})$.
 
Such
an action defines a non-commutative quantum field theory invariant
under the quantum double of $\SU(2)$ (a quantum
deformation of the Poincar\'e group) \cite{PR3,majid,eterawinston,karim,etera}.

Being an invariant function, $F$ can be expanded in group characters: \be F(g)=\sum_{j\in\N/2}
F_j\chi_j(g), \qquad F_0=\int F, \quad F_j=\int dg\,
F(g)\chi_j(g), \ee where the $F_j$'s are the Fourier coefficients
of the Peter-Weyl decomposition in irreducible representations of $\SU(2)$, labelled by $j\in\N/2$.
The kinetic term reads then: \be \kk(g)=1-3F_0^2-\sum_{j\ge 0}\f{F_j^2
}{d_j} \chi_j(g)=\sum_{j\ge 0}
F_j^2\left(1-\f{\chi_j(g)}{d_j}\right)-2F_0^2\,\equiv\,
Q^2(g)-M^2. \ee It is easy to check that $Q^2(g)\ge 0$ with $Q(\id)=0$. We interpret this term as the generalized ``Laplacian" of the theory
while the 0-mode $F_0$ defines the mass $M^2\equiv 2 F_0^2$.

If we choose the simple solution (other choices will give more complicated kinetic terms) \be
F(g)=a+b\chi_1(g), \qquad \int F= a^2+b^2=1, \ee we obtain \be
\kk(g)=\f43(1-a^2)\,\vec{p} \,^2 -2a^2. \ee

\subsection{4d case}
Let us consider a general 4d GFT related to topological BF quantum
field theories, i.e. whose Feynman expansion leads to amplitudes
that can be interpreted as discrete BF path integrals for gauge group ${\cal G}$. This is given by the
following action: \bes S_{4d}&=&\f12\int
[dg]^4\,\phi(g_1,g_2,g_3,g_4)\phi(g_4,g_3,g_2,g_1) \\&&
-\f{\lambda}{5!} \int [dg]^{10}
\phi(g_1,g_2,g_3,g_4)\phi(g_4,g_5,g_6,g_7)\phi(g_7,g_3,g_8,g_9)\phi(g_9,g_6,g_2,g_{10})\phi(g_{10},g_8,g_5,g_1),
\nn \ees where the field is again required to be gauge-invariant,
$\phi(g_1,g_2,g_3,g_4)=\phi(g_1g,g_2g,g_3g,g_4g)$ for any $g\in{\cal G}$. The relevant group for our construction will be $\SO(4,1)$, which requires some
regularization to avoid divergencies due to its non-compact nature.

We generalize to 4d the ``flat solution" ansatz of the 3d group field
theory as \cite{eterawinston}: \be
\phi^{(0)}(g_i)\,\equiv\, {}^3\sqrt{\f{4!}{\lambda}}\int
dg\,\delta(g_1g)F(g_2g)\tlF(g_3g)\delta(g_4g), \ee with $(\int F\tlF)^3=1$. The effective action around such background is \cite{noi}: \bes
S_{eff}[\psi]&=& \f12\int
\psi(g)\psi(g^{-1})\kk(g) \nn\\ && -{}^3\sqrt{\f{\lambda}{4!}}\int
F\int \tlF\,\int\psi(g_1)..\psi(g_3)\,\delta(g_1..g_3) \left[\int
F\int\tlF+\int dh F(hg_3)\tlF(h)\right] \\ &&
-\left({}^3\sqrt{\f{\lambda}{4!}}\right)^2\int F\int \tlF\,\int
\psi(g_1)..\psi(g_4)\,\delta(g_1..g_4) -\f\lambda{5!}\int
\psi(g_1)..\psi(g_5)\,\delta(g_1..g_5), \nn \ees with the
kinetic operator given by: \be \kk(g)\,=\,\left[1-2\left(\int F
\int \tlF\right)^2\int F\tlF -2 \int F\int \tlF \int dh
F(hg)\tlF(h) \int dh F(h)\tlF(hg)\right]. \ee

A simpler special case of the classical solution above is obtained choosing
$\tF(g)=\delta(g)$ while keeping $F$ arbitrary but with $F(\id)=1$. Calling $c\equiv\int F$, the
effective action becomes: \bes
S_{eff}[\psi]&=& \f12\int
\psi(g)\psi(g^{-1})\left[1-2c^2-2cF(g)F(g^{-1})\right]
-c\left({}^3\sqrt{\f{\lambda}{4!}}\right)\,\int\psi(g_1)..\psi(g_3)\,\delta(g_1..g_3)
\left[c+F(g_3)\right] \nn\\ &&
-c\left({}^3\sqrt{\f{\lambda}{4!}}\right)^2\,\int
\psi(g_1)..\psi(g_4)\,\delta(g_1..g_4) -\f\lambda{5!}\int
\psi(g_1)..\psi(g_5)\,\delta(g_1..g_5). \label{compacteffaction}
\ees

In order to make contact with deformed special relativity, we now specialize this construction to one that gives an effective field
theory based on the momentum group manifold $AN_3$.

We start then, as anticipated, from the group field theory describing $\SO(4,1)$
BF-theory.

From the quantum gravity perspective, there are several reasons of interest in this model: 1) the McDowell-Mansouri formulation (as well as related ones
\cite{artem}) 
defines 4d gravity with cosmological constant as a BF-theory for $\SO(4,1)$ plus a potential
term which breaks the gauge symmetry from $\SO(4,1)$ down to the
Lorentz group $\SO(3,1)$; this suggests to try to define Quantum
Gravity in the spin foam context as a perturbation of a
topological spin foam model for $SO(4,1)$ BF theory. These ideas
could also be implemented directly at the GFT level, and the starting point would necessarily be a GFT for $SO(4,1)$ of
the type we use here. 2) we expect \cite{iogft,laurentgft} any
classical solution of this GFT model to represent quantum De
Sitter space on some given topology, and such configurations
would be physically relevant also in the
non-topological case. 3) the spinfoam/GFT model for $\SO(4,1)$ BF-theory
seems the correct arena to build a spin foam model for 4d
quantum gravity plus particles on De Sitter space \cite{artemkg},
treating them as topological curvature defects
for an $SO(4,1)$ connection, similarly to the 3d case \cite{PR3}.

Following the above procedure we naturally
obtain an effective field theory living on $\SO(4,1)$. We want then to obtain from it an effective theory
on $AN_3^{(c)}$. 
We choose: \be F(g)\,=\, \alpha(v_4(g)+a)\vartheta(g),
\qquad \tF(g)\,=\,\delta(g). \ee The function $v_4$ is defined as
matrix element of $g$ in the fundamental (non-unitary)
five-dimensional representation of $SO(4,1)$, $v_4(g)\,=\,\la
v^{(0)}|g|v^{(0)}\ra$, where $v^{(0)}=(0,0,0,0,1)$ is, as
previously, the vector invariant under the $SO(3,1)$ Lorentz
subgroup. $\vartheta(g)$ is a cut-off function providing a
regularization of $F$, so that it becomes an $L^1$ function. Assuming that
$\vartheta(\id)=1$, we require $\alpha=(a+1)^{-1}$ in order for the normalization condition
to be satisfied.

Then we can derive the effective action around such classical
solutions for 2d field variations: \bes S_{eff}[\psi]&=&
\f12\int
\psi(g)\psi(g^{-1})\left[1-2c^2-\f{2c\vartheta^2(g)(a+v_4(g))^2}{(a+1)^2}\right]
-c\left(\f{\lambda}{4!}\right)^{\f{1}{3}}\int\psi(g_1)..\psi(g_3)\,\delta(g_1..g_3)
\left[c+F(g_3)\right] \nn\\ &&
-c\left(\f{\lambda}{4!}\right)^{\f{2}{3}}\,\int
\psi(g_1)..\psi(g_4)\,\delta(g_1..g_4) -\f\lambda{5!}\int
\psi(g_1)..\psi(g_5)\,\delta(g_1..g_5),
\label{noncompacteffaction} \ees where $c=\int F$. Thus the last
issue to address in order to properly define this action is to
compute the integral of $F$; this can be done, and we refer to \cite{noi} for details.

We recognize the
correct kinetic term for a DSR field theory. However, the
effective matter field is still defined on a
$SO(4,1)$ momentum manifold. The only remaining issue is therefore
to understand the ``localization" process of the field $\psi$ to
$\AN^c_3$.

The kinetic term does not show any dependence on the Lorentz
sector. This
suggests that the $\SO(3,1)$ degrees of freedom are non-dynamical
and that the restriction of the field
$\psi$ to $\AN^c_3$ group elements defines the complete dynamics of the theory. This would be trivially true if not for the
fact that the interaction term depend also on the
Lorentz degrees of freedom. One way to make this manifest is, for example,
to assume that the perturbation field $\psi$ has a product
structure $\psi(g)=\tilde{\psi}(h)\Psi(\Lambda)$. The only contribution to the kinetic term from
the Lorentz sector is a constant multiplicative term
$\int_{SO(3,1)}d\Lambda\Psi(\Lambda)\Psi(\Lambda)$. Therefore we
get an exactly DSR-like and $\ka$-Poincar\'e invariant free field
theory. On the other hand, the vertex term couples Lorentz and
$AN_3$ degrees of freedom; thus the $\ka$-Poincar\'e symmetry is
broken and the pure DSR-like form lost. The above also shows that, if we were to choose the dependence of
the perturbation field on the Lorentz sector to be trivial, i.e.
$\Psi(\Lambda)\equiv 1$, and thus to {\it start} from a
perturbation field defined only on the $AN_3$ subgroup, we would
indeed obtain a DSR field theory, but with an interaction
term that would be more complicated that a simple polinomial
interaction, due to to the integrations over the
Lorentz group. Of course, it would be a possible DSR field
theory nevertheless. Still, because of the form
of the kinetic term, we believe a reduction to the $\AN_3$
sector to happen dynamically, or that a proper canonical analysis would show that the
$\SO(3,1)$ modes are pure gauge and can thus drop from the action altogether. Anyway, notice that a restricted
theory obtained from the above and living on $AN_3^c$ only is
dynamically stable. In fact, if we consider only excitations of
the field in $AN^c_3$, we will never obtain excitations in
$\SO(3,1)$ due to momentum conservation $\delta(g_1..g_n)$ since
$\AN^c_3$ is a subgroup. 

Then, restricting ourselves to group elements $h_i\in\AN^c_3$,
we have the field theory: \bes
S_{final}[\psi]= \f12\int \,
\psi(h)\psi(h^{-1})\left[1-2c^2-2cv_4(h)^2\vartheta(h)^2\right]
-c\left(\f{\lambda}{4!}\right)^{\f{1}{3}}\,\int\psi(h_1)..\psi(h_3)\,\delta(h_1..h_3)
\left[c+v_4(h_3)\vartheta(h_3)\right]\nn &&\\
-c\left(\f{\lambda}{4!}\right)^\f{2}{3}\,\int
\psi(h_1)..\psi(h_4)\,\delta(h_1..h_4) -\f\lambda{5!}\int
\psi(h_1)..\psi(h_5)\,\delta(h_1..h_5)\;\;\;\;\;\;\;\;\;\;\;\;\;\;\;\;\;\;\;\;\;\;\;&,& \label{finaldsraction} \nn
\ees with implicit left-invariant measure on $\AN^c_3$. We have thus derived
a DSR scalar field theory with a
$\ka$-deformed Poincar\'e symmetry from the GFT for $\SO(4,1)$ topological
BF-theory.

For other possible strategies leading to the same result, and for more details on the above one, see \cite{noi}.  In particular, notice that we could have directly started from a BF-like GFT action for the group $\AN_3$. Following the same procedure, and with appropriate regularization, we would have obtained easily a DSR field theory of the type we want. What would be less clear, in this case, and this is why we have not focused on this simplified setting, is the link between the initial theory and known classical or quantum formulations of gravity.

\section{Conclusions}
In this contribution, we have briefly reviewed recent results concerning the relation between non-commutative matter field theories, characterized by a configuration space with a Lie algebra structure and a momentum space being a group manifold, and a symmetry group given by quantum deformations of the Poincar\'e group, and group field theories, which are candidate models for the microscopic description of quantum spacetime and of its dynamics, merging the insights of canonical loop quantum gravity, covariant spin foam models and simplicial gravity approaches. In particular, we have seen how the former emerge naturally from the later, in a way reminiscent of the emergence of  field theories for quasi-particles from hydrodynamics in condensed matter theory and analog gravity models. We have motivated this line of research as a  step forward in a research programme trying to bridge the gap between group field theories as fundamental descriptions of a quantum, discrete spacetime at the Planck scale, and the continuum spacetime physics of gravity and  matter at macroscopic scales, using ideas and methods from quantum and statistical field theory and condensed matter physics, applied directly at the GFT level. To be sure: 1) there are many technical and physical aspects of the procedure used to derive non-commutative matter field theories from GFTs that need further analysis; 2) one important difference with respect to analog gravity models should be understood: while quasi-particle dynamics on effective metrics is there obtained from the hydrodynamics of the fundamental system, in our GFT context we have applied a similar procedure directly to the fundamental microscopic field theory; one could speculate that this is why we have obtained {\it non-commutative} field theories for the emergent matter, as opposed to ordinary field theries; 3) the same results have motivations and relevance that are independent from the perspective we have adopted here to present them. These relate to formal aspects of the relation between non-commutative geometry and group field theory, and to the importance of these non-commutative models (and of the general DSR idea) to Quantum Gravity phenomenology \cite{noi}. For all this, we refer to the literature \cite{etera,majid,FKG,QGPhen}. 
However, we believe that these interesting results fit well in the more general scheme we have outlined, on the one hand supporting the picture of GFTs as fundamental theories of quantum gravity, rather than just auxiliary mathematical tools, and on the other hand acquiring, from this more general perspective, an additional reason of interest.

\ack We thank Florian
Girelli and Etera Livine, for the enjoyable collaboration, and for many discussions on
non-commutative field theories, DSR and group field theory, and the organizers of D.I.C.E. for a stimulating conference
and for the opportunity to present our work and ideas.

\end{document}